\documentclass[useAMS,usenatbib,usegraphicx]{mn2e}

\title[Luminosity -- Mass Relation for BCGs]{The luminosity -- halo-mass relation for brightest cluster galaxies}
\author[Brough et al.]{S. Brough$^{1}$\thanks{E-mail:
sbrough@astro.swin.edu.au}, W.~J.~Couch$^{1}$, C.~A.~Collins$^{2}$, T.~Jarrett$^{3}$, D.~J.~Burke$^4$, R.~G.~Mann$^{5}$
\\$^{1}$Centre for Astrophysics and Supercomputing, Swinburne University of Technology, Hawthorn, VIC 3122, Australia
\\$^{2}$Astrophysics Research Institute, Liverpool John Moores University, Egerton Wharf, Birkenhead, CH41 1LD, UK
\\$^{3}$Infrared Processing and Analysis Center, California Institute of Technology, 770 South Wilson Avenue, Pasadena, CA 91125, USA
\\$^{4}$Harvard-Smithsonian Center for Astrophysics, 60 Garden Street, Cambridge, MA 02138, USA
\\$^{5}$SUPA, Institute for Astronomy, University of Edinburgh, Royal Observatory, Blackford Hill, Edinburgh, EH9 3NJ, UK
}

\begin{document}

\date{Accepted... Received...; in original form 2007}

\pagerange{\pageref{firstpage}--\pageref{lastpage}} \pubyear{2007}

\maketitle

\label{firstpage}

\begin{abstract}
We examine the central-galaxy luminosity -- host-halo mass relation
for 54 Brightest Group Galaxies (BGGs) and 92 Brightest Cluster
Galaxies (BCGs) at $z<0.1$ and present the first measurement of this
relation for a sample of known BCGs at $0.1<z<0.8$
($\bar{z}\sim0.3$). At $z<0.1$ we find $L_K\propto
M_{200}^{0.24\pm0.08}$ for the BCGs and the early-type BGGs in groups
with extended X-ray emission and $L_K\propto M_{200}^{0.11\pm0.10}$
for the BCGs alone.  At $0.1<z<0.8$ we find $L_K\propto
M_{200}^{0.28\pm0.11}$.  We conclude that there is no evidence for
evolution in this relationship between $z<0.1$ and $z<0.8$: BCG growth
appears to still be limited by the timescale for dynamical friction
at these earlier times, 
not proceeding according to the predictions of current semi-analytic
models.
\end{abstract}


\section{Introduction}
\label{intro}
Brightest cluster galaxies (BCGs) are the most massive galaxies known
in the Universe and are observed to have remarkably small dispersion
in their absolute magnitudes compared to other early-type cluster
galaxies (e.g. \citealt{aragon-salamanca98,cm98}).  Their position at
the centre of clusters suggests their evolution and their unique
properties are linked to their special environment
(e.g. \citealt{edge91,vonderlinden06}).  However, it is still not
clear exactly which processes drive their evolution.

Of central importance in this context is the mass assembly history of
these most massive galaxies and how that depends on their environment.
Semi-analytic models based on merger trees from the Millennium N-body
simulation predict that, while the stars in BCGs form at
high-redshift, the mass of the galaxy only assembles recently
(doubling in mass since $z\sim1$; \citealt{delucia06}), with the most
massive galaxies in the most massive clusters having undergone the
most mass assembly.  Observations of BCGs agree that their properties
are consistent with having undergone mergers
(e.g. \citealt{boylan-kolchin06,bernardi07}) and BCGs in the most
massive clusters do appear to have undergone more stellar mass
evolution than those in less massive clusters, as shown by their
surface brightness profiles \citep{brough05}.  However, that evolution
must have predominantly occurred before $z\sim1$ as shown by the
uniformity in their absolute magnitudes
\citep{brough02}, and their steep metallicity gradients
\citep{brough07}.

An important clue in this context lies in the central galaxy
luminosity--host-halo mass (L$_{\rm c}$M$_{\rm h}$) relation, which
shows a scaling between BCG mass and that of its host halo.
\cite{lin04} found that $L_K\propto M_{200}^{0.26\pm0.04}$ for 93 BCGs at 
$z<0.1$, \cite{popesso07} found $L_r\propto M_{200}^{0.33\pm0.04}$ for
217 BCGs at $0.01<z<0.25$ in the Sloan Digital Sky Survey (SDSS), and
\cite{hansen07} found $\langle L_i\rangle \propto
M_{200}^{0.30\pm0.01}$ for the SDSS MaxBCG cluster sample
($0.1<z<0.3$).
Studies of the L$_{\rm c}$M$_{\rm h}$-relation have not ventured
beyond $z\sim0.3$, so it is not clear whether this relationship
evolves with cosmic time.

The L$_{\rm c}$M$_{\rm h}$-relation is also a natural output from
analyses of cosmological simulations: Halo Occupation Distribution
functions (HOD; e.g. \citealt{vandenbosch05,zheng07}) and Conditional
Luminosity Functions (CLF; e.g. Cooray \& Milosavljevi\'{c} 2005;
henceforth CM05).  These statistically assign numbers of galaxies or
the luminosity distribution of galaxies in a given dark matter halo,
distinguishing between central and satellite galaxies and enabling a
direct comparison with observations.  A related approach is the
application of a direct relationship between the mass of a dark matter
halo or sub-halo and the observed galaxy luminosity
(e.g. \citealt{vale06,vale07}; henceforth VO06, VO07).  CM05 predict
$\langle L\rangle \propto M_{200}^{<0.3}$ above halo masses of
$4\times10^{13}h^{-1}M_{\odot}$ as the timescale on which dynamical
friction causes satellite galaxies to fall in to the centre exceeds
the age of the host system halo.
VO06 predict $\langle L_{K} \rangle \propto M_{100}^{0.28}$ as a
result of the hierarchical formation of structure in the Universe,
with a prescription for sub-halo mass loss.  We interpret the L$_{\rm
c}$M$_{\rm h}$-relation in this paradigm: that it results from the BCG
growing in step with its host halo.

\cite{yang05} analysed the HOD of galaxy groups in the 2-degree 
Field Galaxy Redshift Survey ($0.01<z<0.2$) and found $\langle
L_{cen,b_J}\rangle \propto M_h^{0.25}$.  More recently, \cite{yang07}
determined the CLF of groups ($0.01<z<0.2$) in the SDSS and found
$\langle L_{cen,r}\rangle \propto M_h^{0.17}$.  At higher redshifts,
\cite{zheng07} analysed the HOD using the two-point correlation function
of the DEEP2 galaxy redshift survey (${\bar z}\sim1$) in comparison to
the SDSS (${\bar z}\sim0$) and determined the 
L$_{\rm c}$M$_{\rm h}$-relation and its evolution.  They found
the VO06 model to be a good fit to their data at $z\sim0$ and $z\sim1$
suggesting that this relationship has not evolved.  \cite{white07}
found $\langle L_{cen,B}\rangle \propto M_h^{0.36}$ at $z\sim0.5$ from
their HOD analysis of $\sim2L_{\star}$ red galaxies in the NOAO Deep
Wide Field Survey.  These studies, however, do not directly identify
central galaxies.

In this Letter, we examine the central-galaxy luminosity -- host halo
mass relation for a sample of known BCGs in groups and clusters at
$z<0.1$ and, for the first time, the relationship for known BCGs in
clusters at redshifts $0.1<z<0.8$.  This will enable us to directly
determine the evolution in this relationship for known BCGs and to
compare to the HOD analyses at these redshifts. It will also enable us
to examine the \cite{delucia06} prediction of these galaxies having
doubled in mass since $z\sim1$.

In Section~\ref{Data} we introduce our sample and then present our
comparison of BGG and BCG luminosities with host system mass in
Section~\ref{Results}.  We discuss the implications of our results in
Section~\ref{Discussion} and draw our conclusions in
Section~\ref{Conclusions}.  Throughout this paper we assume $H_0=70$
km s$^{-1}$ Mpc$^{-1}$, $\Omega_m=0.3$ and $\Omega_{\Lambda}=0.7$.

\section{Data}
\label{Data}

The BCG sample is from \cite{brough02} and contains 92 BCGs at
$z\leq0.1$ and 63 at $0.1<z<0.8~(\bar{z}\sim0.3)$.  These are selected
as the brightest galaxies closest to the X-ray centroid of their host
cluster and are all early-type galaxies.

In \cite{brough02}, $K$-band magnitudes were measured in an aperture of
radius $12.5 h^{-1}$ kpc.  The apparent magnitudes were corrected to
absolute magnitudes using
K- and passive evolution corrections 
\citep{yoshii88}
produced by the GISSEL96 stellar population synthesis code
\citep{bruzual93}. 
The galaxies were assumed to be 10\,Gyrs old, to have formed in an
instantaneous burst, and to have evolved passively since $z\sim2$.  The
assumptions of age and redshift of formation have negligible effect on
the correction value. The magnitudes were also corrected for galactic
absorption (typical values $A_K\sim0.02$ mag) using the maps of
\cite{schlegel98}.  The uncertainties in these magnitudes are $\sim0.04$ mag.

Cluster X-ray luminosity is directly proportional to the square of the
electron density of the intra-cluster medium and provides a measure of
the host system mass.  X-ray luminosities for the low-redshift
clusters were measured from the {\it ROSAT} All Sky Survey (RASS),
paired with co-ordinates from Abell et al. (1989) and
\citealt{lynam99}.  The X-ray luminosities for the high-redshift
clusters were obtained from the Einstein Medium Sensitivity Survey
(EMSS; \citealt{gioia94}) and Serendipitous High-Redshift Archival
Cluster {\it ROSAT} catalogues \citep{burke03}.  In \cite{brough02}
the RASS passband ($0.1-2.4$ kev) data were transformed to the EMSS
passband ($0.3-3.5$ keV).  Here we use the same transformation to
place all these values in the RASS passband, i.e. $L_X (0.1-2.4$
keV)$=L_X (0.3-3.5$ keV)$ / 1.08$.  The host clusters cover a wide
range in X-ray luminosity: $5\times10^{42}< L_X<5\times10^{44}$
($h^{-1}$ erg s$^{-1})$ at redshifts $z<0.1$ and $6\times10^{42}<
L_X<1\times10^{45}$ ($h^{-1}$ erg s$^{-1})$ at redshifts $0.1<z<0.8$.

The Group Evolution Multiwavelength Survey (GEMS;
\citealt{osmond04,forbes06}) is a heterogeneous sample of 60
groups selected from optically-selected group catalogues observed by
the {\it ROSAT} X-ray satellite to enable X-ray classification.
Removing the 6 groups for which
\cite{osmond04} were unable to find at least 4 galaxies associated
with the original optical position, leaves a sample of 54 groups with
$2\times10^{40}<L_X<4\times10^{43}$ ($h^{-1}$ erg s$^{-1})$.  Of these
54 groups, 35 have extended intra-group X-ray emission, 13 have X-ray
emission only associated with the central galaxy and 6 are undetected
(at $3\sigma$ above background) in X-rays.  \cite{osmond04} also
identified the BGG and published the galaxies' morphologies.

All the GEMS groups are covered by the 2 Micron All Sky Survey
Extended Source Catalogue (2MASS; \citealt{jarrett00}).  Our
photometric reductions were performed to be as consistent as possible
with \cite{brough02}.  All 54 groups are at redshifts $z<0.03$
($\bar{z}\sim0.01$) and for a large proportion of the sample the
circular aperture photometry in 2MASS does not extend to large enough
radii to measure an aperture of radius $12.5 h^{-1}$ kpc (the maximum
possible radius $70^{\prime\prime}=14 h^{-1}$ kpc at $z\sim0.01$).  The
aperture magnitudes for this sample were therefore measured manually
from the 2MASS scans and corrected to absolute magnitudes following
\cite{brough02}.   They have photometric uncertainties of $\sim0.04$ mag.

The group X-ray luminosities are taken from the GEMS group catalogue.
These are bolometric luminosities, extrapolated to the radius
corresponding to an overdensity of 500 times the critical density
($r_{500}$), with uncertainties of log$L_X\sim0.05$.  \cite{brough07}
showed that the GEMS group X-ray luminosities are consistent with
those measured by the {\it ROSAT}-ESO Flux Limited X-ray galaxy
cluster survey (REFLEX;
\citealt{bohringer04}) extrapolated to 12 times the core radius of the
cluster, with an uncertainty of log$L_X=0.05$.  We repeated our
low-redshift group and cluster analysis using REFLEX X-ray
luminosities for the 40 clusters that are in common with this work and
found that our results are unchanged.  We therefore present this work
using the full low-redshift sample of 92 clusters.

We convert the group X-ray luminosities to the mass within an
overdensity of 200 times the critical density, M$_{200}$, using the
$L_{X}(r_{500},{\rm Bol})-$M$_{200}^{1.13\pm0.27}$ relation measured
for 15 GEMS groups from \cite{brough06}.  We then use the
$L_X(0.1-2.4$~keV$)-$M$_{200}^{1.57\pm0.08}$ relation for clusters
from \cite{reiprich02} for our clusters.  We estimated the
uncertainties in our mass measurements by performing a Monte-Carlo
analysis of how the uncertainties in the $L_{X}$ measurements
propagate through to our mass estimates given the uncertainty in the
slope of the $L_{X}-$M$_{200}$ relation used for our conversion.
These were found to be of order 0.08 in log M$_{200}$ for the groups
and 0.02 for the clusters.  
\cite{ettori04} show that the \cite{reiprich02}
relationship does not evolve significantly over the redshift range
analysed here, we therefore use the cluster
$L_X(0.1-2.4$~keV$)-$M$_{200}$ relationship for our higher redshift
sample.

\subsection{Aperture Magnitudes}

\begin{figure}
\begin{center}

    \resizebox{16pc}{!}{
     \rotatebox{-90}{
	\includegraphics{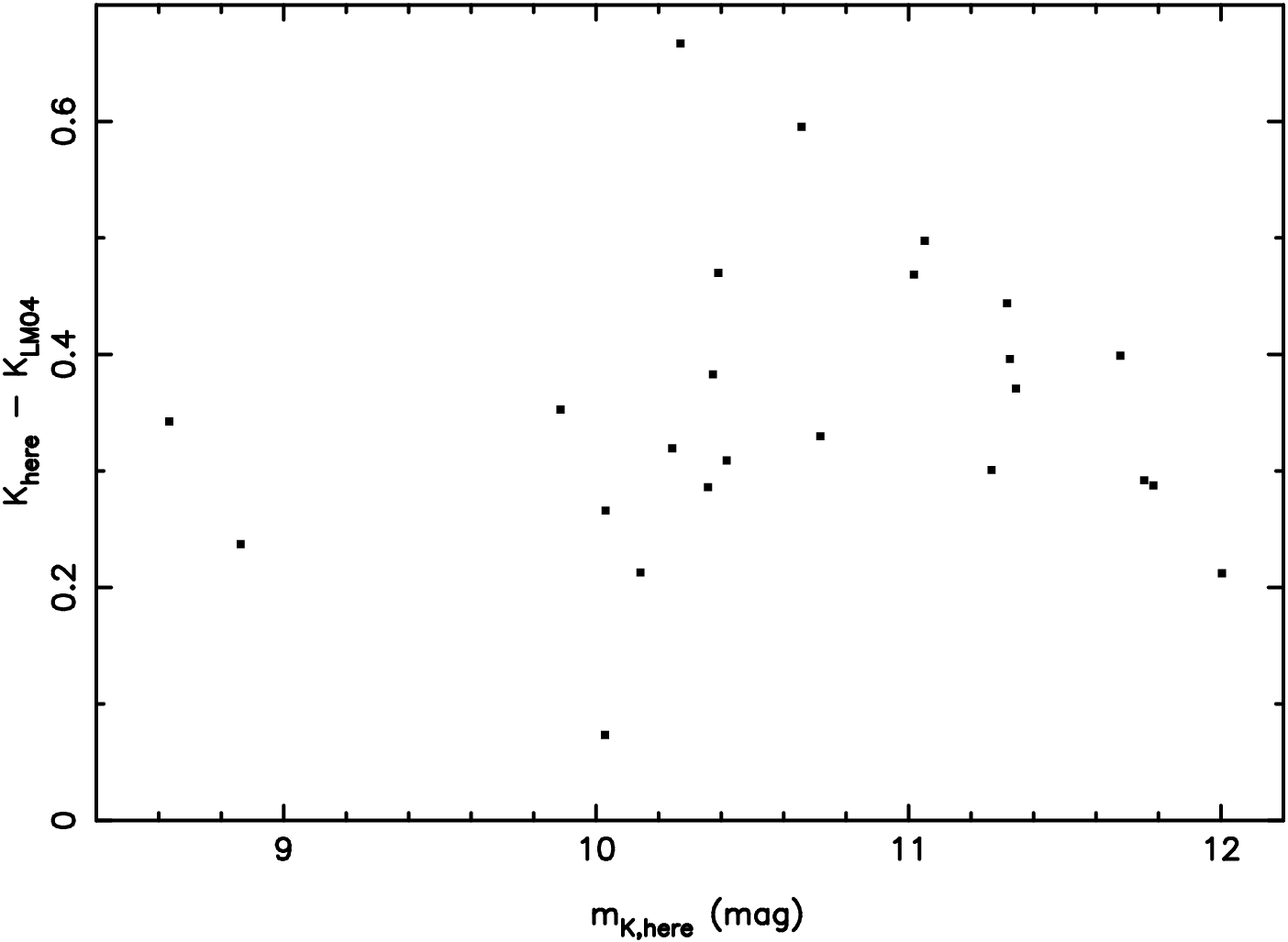}
	\includegraphics{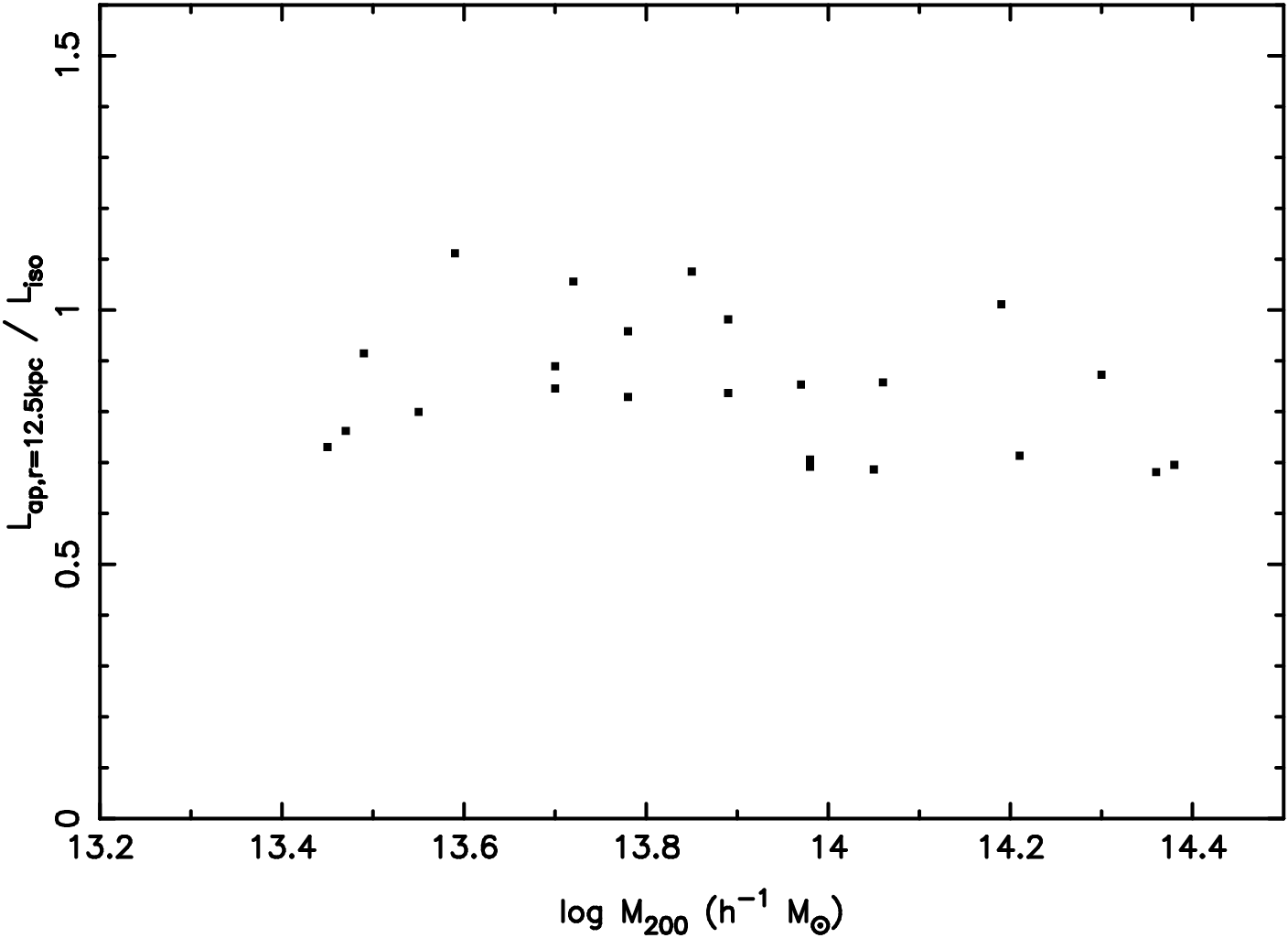}
    }
}
  \end{center}
\caption {{\it Upper Panel}: Comparing our apparent aperture magnitudes with the 
apparent isophotal magnitudes from Lin \& Mohr (2004).  We find an
offset of $\Delta m_K=0.35\pm0.03$ mag for the 24 galaxies in common.
{\it Lower Panel}: Comparing model aperture and isophotal luminosities
with host system mass.  }
\label{comp_mag}
\end{figure}

In their BCG study, \cite{lin04} make the case that aperture
magnitudes are not sensitive to the mass growth of BCGs and instead
use 2MASS $K$-band isophotal magnitudes (measured within the 20th
magnitude isophote).  We test whether this affects our data in several 
different ways:

We compared the apparent magnitudes for the 24 galaxies in common with
\cite{lin04} and found an offset of $\Delta m_K=0.35\pm0.03$ mag
resulting from the use of the different measurements
(Figure~\ref{comp_mag}).  This offset is not correlated with the
magnitudes or with cluster mass.

We re-measure the magnitudes in a larger aperture of $25 h^{-1}$kpc
for 4 BCGs at the same mean luminosity in high-mass ($\bar{M}\sim
4\times10^{14}h^{-1}M_{\odot}$), and in low-mass clusters
($\bar{M}\sim1\times10^{14}h^{-1}M_{\odot}$).  If we were subject to
aperture effects then the magnitudes of the galaxies in the low-mass
clusters would increase less than those in the high-mass clusters.
However, the increases are consistent: $\Delta M_K
(\rm{low~mass})=0.39\pm0.02$, $\Delta M_K
(\rm{high~mass})=0.44\pm0.04$.

Finally, we use the $R$-band surface brightness profiles from
\cite{graham96} of 24 galaxies in common with this work.  We then 
predicted (assuming $R-K=2.6$) aperture luminosities within $r=12.5
h^{-1}$ kpc and isophotal luminosities within $K=20$ magarcsec$^{-2}$.
The lower panel of Figure~\ref{comp_mag} shows that aperture
luminosities follow the same relationship as isophotal
luminosities with cluster mass.

We therefore conclude that our results are robust to any possible
aperture effects.

\section{Results}
\label{Results}


\begin{figure}
\begin{center}

    \resizebox{20pc}{!}{
     \rotatebox{-90}{
	\includegraphics{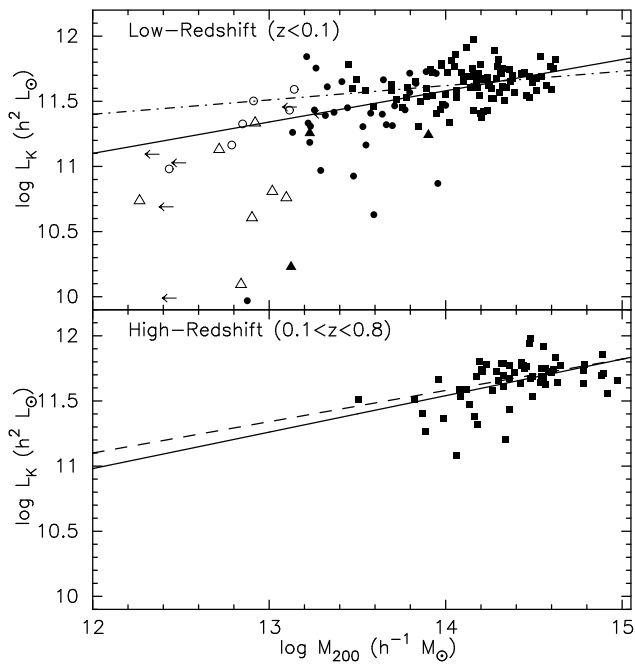}
    }
}
  \end{center}
\caption {The relationship of BGG and BCG aperture $K$-band luminosities 
with host system mass.  The upper panel shows the relationship for the
BGGs and the BCGs in low-redshift clusters ($z\leq0.1$).  In this
panel the points indicate the BCGs (filled squares) and the properties
of the BGGs (circles denote early-type galaxies, triangles denote
late-types) and their host groups (filled points indicate extended
X-ray emission, open points indicate galaxy halo emission and arrows
indicate upper limits for the X-ray undetected groups).
The solid line indicates the $L_K\propto M_{200}^{0.24\pm0.08}$
relation for the (eBGG+BCG sample), while the dot-dashed line
indicates the $L_K\propto M_{200}^{0.11\pm0.10}$ relation for the BCGs
alone.  The lower panel shows the relationship for the BCGs in
high-redshift clusters ($0.1<z<0.8$). The solid line indicates the
best-fit $L_K\propto M_{200}^{0.28\pm0.11}$ relation for these
high-redshift data and the dashed line indicates the low-redshift
eBGG+BCG sample for direct comparison.}
\label{L_M_both}
\end{figure}

In the upper panel of Figure~\ref{L_M_both} we plot the BCG ($z<0.1$)
and BGG aperture $K$-band luminosities versus their host system mass.
The galaxies in the lower-mass systems are fainter and,
morphologically, more likely to be late-type galaxies (Sa to Irr) in
systems without extended intra-group X-rays. The dispersion in the
central galaxy luminosities decreases as the host halo mass increases,
as also observed by \cite{zheng07}.  They suggest that this is a
result of a broader distribution of major star formation epochs in
lower mass halos.  This is consistent with our observation that many
of the galaxies in the low-mass halos are late-types.

Fitting a relationship to these data we find $L_K\propto
M_{200}^{0.46\pm0.06}$ for all BGGs and BCGs.  This is steeper than
has previously been observed.  However, given that the extended X-ray
emission characteristic of the potential well of a bound system is not
detected in the less-massive groups, it is more reliable to determine
this relationship from the early-type BGGs in groups with extended
X-ray emission and BCGs (eBGG+BCG sample).  For these we find a
relationship $L_K\propto M_{200}^{0.24\pm0.08}$.  This is consistent
with values from the literature
(e.g. \citealt{lin04,popesso07,hansen07}). We note that the
uncertainties in these data have insignificant effect on the fitted
relations.
Analysing the BCGs alone yields a shallower relation: $L_K\propto
M_{200}^{0.11\pm0.10}$.  This relationship is consistent with no
increase of BCG luminosity with increasing cluster mass.  However, due
to the scatter in the data it is also consistent with the relationship
for the eBGG+BCG sample and with those relationships in the
literature.

Whether the L$_{\rm c}$M$_{\rm h}$-relation evolves with redshift is
also interesting.  We examine the 63 BCGs at higher redshift
($0.1<z<0.8$) in the lower panel of Figure~\ref{L_M_both}.  The
best-fitting relationship to these data is $L_K\propto
M_{200}^{0.28\pm0.11}$.  This is entirely consistent with the
low-redshift relations.

\begin{figure}
\begin{center}

    \resizebox{16pc}{!}{
     \rotatebox{-90}{
	\includegraphics{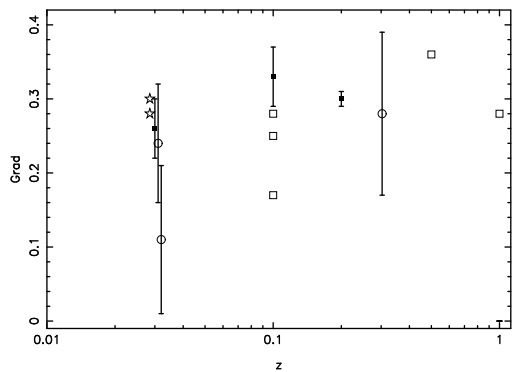}
    }
}
  \end{center}
\caption {Comparing our results with gradients from L$_{\rm c}$M$_{\rm h}$-relation studies 
in the literature with redshift.  Stars indicate the predictions from
CM05 and VO06, filled squares indicate previous BCG studies, open
squares indicate previous HOD studies and open circles indicate the
results presented here. }
\label{comp_studies}
\end{figure}

In Figure~\ref{comp_studies} we show how our results compare to those
measured in the literature, and to the predictions of CM05 and VO06.
To add another high-redshift ($z\sim 1$) point, we assume that because
the \cite{zheng07} data fit the VO06 model, they can be represented by
$L\propto M^{0.28}$.  There is no appreciable offset between the
relationship measured from HOD analyses or directly from BCG samples.
These suggest that there is no evolution in this relationship between
$z\sim0.01$ and $z\sim1$ .  Our data are also consistent with the
predictions of both CM05 and VO06.


\section{Discussion}
\label{Discussion}

BCGs are often thought to be the most conspicuous examples of central
cluster galaxies undergoing rapid growth through merging.
These data suggest a picture in which the near-infrared luminosity
($\sim$stellar mass) of early-type BCGs evolves steadily with the
growth of their host cluster.  In observing the L$_{\rm c}$M$_{\rm
h}$-relation to have a slope $\sim0.3$ at higher redshift, BCG growth
appears to still be in the `dynamical friction time-scale-limited'
regime (as per CM05) at these earlier times.  Hence, BCG-halo mass
growth is controlled by this restriction over the cosmic time interval
studied here, and is not consistent with BCGs doubling in mass from
$z\sim1$ to $z\sim0$ as suggested by \cite{delucia06}.  These
observations are more consistent with BCGs accreting $10-20$ per cent
in stellar mass as their cluster doubles in mass.

This conclusion is consistent with the observation that the
position of BCGs on scaling relations such as the fundamental plane is
a result of having undergone at least 1 major merger event
(e.g. \citealt{boylan-kolchin06,bernardi07}). These major merger
events have been directly observed:
\cite{yamada02} observed a nearly
equal-mass BCG merger in a $\sim2\times10^{14}h^{-1}M_{\odot}$ cluster
at $z\sim1.26$.
\cite{rines07} have observed a 
plume around a
BCG in a $\sim1\times10^{14}h^{-1}M_{\odot}$ cluster at
$z\sim0.4$. Both of these systems lie within the scatter of our data
on the L$_{\rm c}$M$_{\rm h}$-relation.

Observations of the mass assembly of other massive early-type galaxy
populations are contradictory. Some claim evidence for no evolution
beyond passive for these galaxies (e.g. \citealt{cimatti06,bundy06}),
whilst others claim stellar mass evolution of a factor of 2 since
$z\sim1$ (e.g. \citealt{vandokkum05,bell06}).  In between these extremes
are studies more consistent with this work, finding evidence for
growth of $\leq50$ per cent since $z\sim1$
(e.g. \citealt{glazebrook04,wake06,brown06,white07,masjedi07,mcintosh07}).


From a theoretical perspective, we have demonstrated that our data are
consistent with the predictions of two HOD models at $z\sim0.1$.
However, these models do not make predictions for higher redshifts.
Examining the evolution of BCGs in a $\Lambda$-Cold Dark Matter
Universe, the semi-analytical model of \cite{delucia06} predicts that
$\sim50$ per cent of the mass of BCGs is assembled since $z\sim1$,
with an equivalent growth in the host halo over this timescale.  This
is inconsistent with our observations.  However,
\cite{almeida07} have analysed the stellar mass evolution of large red
galaxies between $z\sim0.24$ and $z\sim0.5$ in the Durham
semi-analytic models, and find that, on average, large red galaxies
assemble $\sim25$ per cent of their mass since $z\sim1$.

The results presented in this letter are clearly consistent with the
picture emerging from studies of the evolution of massive early-type
galaxies and semi-analytic models of the evolution of these massive
galaxies, that massive galaxies have increased in stellar mass by
$\leq50$ per cent since $z\sim1$.


\section{Conclusions}
\label{Conclusions}
We have examined the BGG and BCG luminosity -- host halo mass relation
at: $z<0.1$ and, at higher redshifts than previously measured:
$0.1<z<0.8$.  We conclude that:

\begin{itemize}

\item BCGs follow a relationship with the mass of their host cluster. 
This relationship does not evolve from $z\sim1$ to $z\sim0$.

\item Our data are consistent with the predictions of two illustrative 
HOD models for the relationship of central galaxy luminosity with host
halo mass. 

\item The BCG -- halo-mass relation appears to be controlled by the timescale 
for dynamical friction even at $0.1<z<0.8$.

\end{itemize}

\section*{Acknowledgements}

We thank the referee for instructive comments.  We also thank
M.J.I. Brown, C. Blake, A. Arag\'{o}n-Salamanca and A. Edge for
helpful discussions. SB and WJC acknowledge the support of the
Australian Research Council.  DJB acknowledges the support of NASA
contract NAS8-39073.  This publication makes use of data products from
the Two Micron All Sky Survey (2MASS) which is a joint project of the
University of Massachusetts and the Infrared Processing and Analysis
Center/California Institute of Technology, funded by the National
Aeronautics and Space Administration and the National Science
Foundation.

\bsp

\label{lastpage}

\end{document}